%% file: main.tex
\documentclass[runningheads]{llncs}

\input{preamble/packages}
\input{preamble/definitions}

\input{preamble/authors}

\begin{document}

\title{Revisiting Language Models in \\ Neural News Recommender Systems}

\titlerunning{Revisiting Language Models in Neural News Recommender Systems}

\maketitle

\input{abstract}

\acresetall

\input{sections/1-introduction}
\input{sections/2-background}
\input{sections/3-reproducibility}

\input{sections/4-experiments}
\input{sections/5-limitations}
\input{sections/6-conclusion}

\subsection*{Acknowledgments}
This research was (partially) supported by the Dutch Research Council (NWO), under project numbers 024.004.022, NWA.1389.20.\-183, and KICH3.LTP.20.006, and the European Union's Horizon Europe program under grant agreement No 101070212.

All content represents the opinion of the authors, which is not necessarily shared or endorsed by their respective employers and/or sponsors.

\bibliographystyle{splncs04nat} %
\bibliography{references}

\end{document}

%% file: preamble/packages.tex
\usepackage{physics}
\usepackage{amsmath}
\usepackage{amssymb}
\usepackage{xcolor}
\usepackage{acronym}
\usepackage{graphicx}
\usepackage[skip=0pt]{caption}
\usepackage{subcaption} 
\usepackage{tablefootnote}
\usepackage{tabularx}
\usepackage{threeparttable}
\usepackage{booktabs}
\usepackage{multirow}
\usepackage{siunitx}
\usepackage[numbers,sort&compress]{natbib}  
\usepackage[inline]{enumitem}
\usepackage[para]{footmisc}
\usepackage{xspace}
\usepackage{wrapfig,lipsum,booktabs}
\usepackage{bm}
\usepackage{pifont}

\usepackage[T1]{fontenc}
\usepackage{acronym}
\usepackage{graphicx}

\usepackage{hyperref}
\hypersetup{
  colorlinks   = true,    %
  urlcolor     = blue,    %
  linkcolor    = blue,    %
  citecolor    = red      %
}

%% file: preamble/definitions.tex
\newcommand{\header}[1]{\vspace*{1mm}\noindent\textbf{#1.}}

\definecolor{yuyueblue}{HTML}{64afdc}
\definecolor{tablered}{HTML}{000000}
\definecolor{tableblue}{HTML}{000000}

\newcommand{\ie}{\emph{i.e.,}\xspace}

\newcommand{\eg}{\emph{e.g.,}\xspace}

\newcommand{\ignore}[1]{}

\makeatletter
\def\@fnsymbol#1{\ensuremath{\ifcase#1\or \dagger\or \ddagger\or
   \mathsection\or \mathparagraph\or \|\or **\or \dagger\dagger
   \or \ddagger\ddagger \else\@ctrerr\fi}}
    \makeatother
    
\acrodef{RS}{recommender system}
\acrodef{LM}{language model}
\acrodef{LLM}{large language model}
\acrodef{NLM}{neural language model}
\acrodef{SLM}{shallow language model}
\acrodef{PLM}{pre-trained language model}
\acrodef{NLP}{natural language processing}
\acrodef{AI}{artificial intelligence}
\acrodef{IR}{information retrieval}

%% file: preamble/authors.tex
\author{
Yuyue Zhao\inst{1}\orcidID{0000-0002-5298-0309} \and
Jin Huang\inst{1,2}\thanks{Corresponding author.}\orcidID{0000-0001-9273-9037} \and
David Vos\inst{1}\orcidID{0009-0003-8925-1585} \and 
Maarten de Rijke\inst{1}\orcidID{0000-0002-1086-0202} 
}

\institute{
University of Amsterdam \and Delft University of Technology 
\\
\email{y.zhao3@uva.nl, j.huang2@uva.nl, d.j.a.vos@uva.nl, m.derijke@uva.nl}
} 

\authorrunning{Y. Zhao et al.}

%% file: abstract.tex
\begin{abstract}
    Neural news \acp{RS} have integrated \acp{LM} to encode news articles with rich textual information into representations, thereby improving the recommendation process.
    Most studies suggest that (i)~news \acp{RS} achieve better performance with larger \acp{PLM} than \acp{SLM}, and (ii)~that \acp{LLM} outperform \acp{PLM}.
    However, other studies indicate that \acp{PLM} sometimes lead to worse performance than \acp{SLM}.
    Thus, it remains unclear whether using larger \acp{LM} consistently improves the performance of news \acp{RS}.
    In this paper, we revisit, unify, and extend these comparisons of the effectiveness of \acp{LM} in news \acp{RS} using the real-world MIND dataset.
    We find that (i)~larger \acp{LM} do not necessarily translate to better performance in news \acp{RS}, and (ii)~they require stricter fine-tuning hyperparameter selection and greater computational resources
    to achieve optimal recommendation performance than smaller \acp{LM}.
    On the positive side, our experiments show that larger \acp{LM} lead to better recommendation performance for cold-start users: they alleviate dependency on extensive user interaction history and make recommendations more reliant on the news content.
    
    \keywords{News recommendation  \and Language model \and Fine-tuning.}
\end{abstract}

%% file: sections/1-introduction.tex
\section{Introduction}
\label{sec:intro}
News \acp{RS} help deliver relevant news articles to users.
Unlike \acp{RS} in other domains, such as e-commerce and music, that primarily focus on modeling interactions between users and items, news \acp{RS} rely heavily on modeling text-based news articles with rich textual information~\citep{surveyNews19}.
Therefore, \acl{NLP} techniques, particularly methods based on \acp{LM}, are widely used to generate news representations in news \acp{RS}.

Among the early \ac{LM}-based approaches to news representation are \acp{SLM} such as GloVe~\citep{glove}, a model that generates word representations based on corpus co-occurrence statistics.
In news \acp{RS}, GloVe embeddings are used to initialize word embeddings, which are later employed to model news articles and interactions~\citep{nrms,naml,lstur}.
Following progress in language modeling, \acp{PLM} such as BERT~\citep{bert} and RoBERTa~\citep{roberta} have also been integrated into news \acp{RS} to generate embeddings for news articles.
Compared to \acp{SLM}, \acp{PLM} are typically larger, featuring complex architectures with more layers, thus contributing to a greater number of parameters.
\Acp{LLM} such as Llama~\citep{llama} are also used to enhance news modeling in~\acp{RS} because of their ability to capture context and to generalize~\citep{once24, promptNRSecir}.

Most prior work in news \acp{RS} shows that news \acp{RS} using larger \acp{PLM} outperform those using \acp{SLM}~\citep{plm-nr, speedNRS, miner, unbert, efficientFed, distillnewsbert, fairNRS, tadi23}.
Work using \acp{LLM} has shown better recommendation performance than \acp{PLM}~\citep{once24, promptNRSecir}.
But the findings are not consistent: some work reports that \acp{PLM} sometimes perform worse than \acp{SLM} in news \acp{RS}~\citep{newsreclib, multimind}.
Are larger models worth the additional computational resources?
We examine (i) whether larger \acp{LM}
truly improve news recommendation performance, and (ii) what
size \acp{LM} (as news encoders) provides a reasonable trade-off between performance and resource consumption.

To answer these questions, we compare the impact of using eight \acp{LM} -- across different \ac{LM} families, i.e., GloVe (\ac{SLM}), BERT and RoBERTa (\acp{PLM}), and  Llama3.1-8B (\ac{LLM}), as well as multiple sizes within the BERT family (tiny, mini, small, medium, and base) -- on the performance of three well-known news recommendation models: NAML~\citep{naml}, NRMS~\citep{nrms}, and LSTUR~\citep{lstur}.
Consistent with widely adopted practices~\citep{once24, promptNRS, mtrec22, promptNRSecir}, our experiments are based on the small version of the real-world MIND dataset~\citep{mindDataset}.

We focus on the following research questions: 
\begin{enumerate}[label=\textbf{RQ\arabic*},leftmargin=*,ref={RQ\arabic*}]
    \item Does using a larger \ac{LM} in news RSs consistently lead to better recommendation accuracy? \label{rq1}
    \item How does fine-tuning \acp{LM} affect the performance of LM-based news RSs?\label{rq2}
\end{enumerate} 
\acp{LM} may enhance the performance of news RSs for cold-start users with limited or no user interaction history by analyzing the textual content of news articles and recommending relevant content based on extracted semantic information~\cite{sanner2023large,wang2024large}. 
Therefore, our third research question concerns the recommendation performance of different LMs for cold-start users:
\begin{enumerate}[label=\textbf{RQ\arabic*},leftmargin=*, resume,ref={RQ\arabic*}]
    \item Do news \acp{RS} based on larger \acp{LM} provide better performance for cold-start users? \label{rq3}
\end{enumerate} 

\noindent%
Larger \acp{LM} in news \acp{RS} do not always lead to improved performance of recommendations.
The performance of LM-based news RSs depends heavily on whether the LMs are fine-tuned.
E.g., without fine-tuning, NRMS using the \ac{SLM} GloVe outperforms NRMS using \acp{PLM} BERT and RoBERTa, and even performs comparable to NRMS using \ac{LLM} Llama. 
Moreover, while larger LMs require more comprehensive fine-tuning, such as searching for the optimal number of fine-tuned layers, they tend to achieve better performance for cold-start users. 
LM-enhanced news encoders alleviate the dependency on user interaction history, making recommendations more reliant on news content itself.

%% file: sections/2-background.tex
\vspace*{-2mm}
\section{Related Work}

\vspace*{-2mm}
\textbf{Selection criteria for related work.}
\label{sec-sourceselection}
We follow the guidelines in \citep{guidelines} to select relevant literature on \acp{LM} as news encoders for news recommendation. Sources are chosen from top venues and journals in the fields of \ac{AI} and \ac{IR}. Papers are included if they 
(i) propose a definition of text modeling in the context of news recommendation, 
(ii) introduce approaches to improve news recommendation performance, or 
(iii) present experimental results comparing the performance of different-sized \acp{LM} as news encoders using the same benchmark.
Papers are excluded if (i) their approaches are not tested on an English news recommendation dataset or (ii) they fall outside the date range of October 2014 (the release of GloVe) to October 2024.
We identified over 200 studies, 24 of which are highly relevant to our work. 
Below, we introduce these studies to provide context for our research.

\header{LMs in news RSs} News content modeling is a crucial component of news \acp{RS}, as news articles contain rich textual information that can be effectively encoded using \acp{LM}~\citep{surveyWuNews}. 
Following the categorization criteria in~\citep{llmsurvey, llmsurveyHOW}, methods using \acp{LM} in news \acp{RS} can be grouped based on the role of the \ac{LM}:
(i) \acp{LM} as news recommenders, which generate candidate news items~\citep{promptNRSecir, gptNRS},
(ii) \acp{LM} as news encoders, which encode news content to support news \acp{RS}~\citep{plm-nr, speedNRS, miner, unbert, efficientFed, distillnewsbert, fairNRS, tadi23, newsreclib}, and
(iii) \acp{LM} as news enhancers, which generate additional textual features that assist news \acp{RS}~\citep{once24, categoryNewsG}.
In this study, we focus on the largest group, where \acp{LM} are used as news encoders to explore the impact of different \acp{LM} in news \acp{RS} in relation to their effectiveness and efficiency.

\header{LMs as news encoders}
Early \ac{LM}-based approaches to news \acp{RS} learn representations on \acp{SLM}, such as GloVe.  
NAML~\citep{naml} uses GloVe embeddings to initialize word representations and employs a word-level and view-level attention mechanism, along with convolutional neural networks, to capture important words for news representation. 
NRMS~\citep{nrms} uses GloVe embeddings for initialization and adopts multi-head self-attention to learn news representation. 
Prior work has argued that such shallow models may not be sufficient to capture the semantic information in news articles, and has explored \acp{PLM} based on the transformer architecture~\citep{attention}, such as BERT, for news modeling. 
E.g., PLM-NR~\citep{plm-nr} uses \acp{PLM} to enhance news representation and observes improvements over \acp{SLM} as a news encoder model. 
MINER~\citep{miner} employs a pre-trained BERT as the news encoder and uses a poly-attention mechanism to extract multiple aspect interest vectors for users.
More recently, \acp{LLM} have been explored for news modeling. 
ONCE~\citep{once24} uses both open- and closed-source \acp{LLM} to enrich training data and enhance content representation. 
\citet{categoryNewsG} improve news recommendations by using \acp{LLM} to generate category descriptions. 
PGNR~\citep{promptNRSecir} employs \acp{LLM} to frame news recommendation as a text-to-text generation task, performing recommendation through generation.

According to the studies listed above, transitioning from \acp{SLM} to \acp{PLM} and then to \acp{LLM} results in clear improvements in news recommendation performance.
However, some studies report different findings.
NewsRecLib~\citep{newsreclib}, a widely used news recommendation benchmark, reports that NAML and LSTUR, which originally used GloVe, performs worse when GloVe is replaced by the \ac{PLM} BERT.
Additionally, xMIND~\citep{multimind}, a publicly available multilingual benchmark for news recommendation, indicates that NAML using \ac{PLM}-based embeddings performs worse than the version using randomly-initialized embeddings.
These contradictory results highlight the need for further investigation into the effectiveness of \acp{LM} in news recommendation, which motivates this study.

%% file: sections/3-reproducibility.tex
\vspace*{-2mm}
\section{Reproducibility Methodology}

\vspace*{-2mm}
\subsection{Problem formulation}
\vspace*{-1mm}
Let $\mathcal{V}$ represent the set of news articles, where each news article $v \in \mathcal{V}$ consists of its textual feature $f_t(v)$ (e.g., title or abstract) and other features $f_d(v)$ (e.g., news categories or subcategories). 
Let $\mathcal{U}$ represent the set of users.
Each user $u \in \mathcal{U}$ has a click history $H_u = \{v^h_1, v^h_2, \ldots, v^h_n\}$ in chronological order, denoting the sequence of $n$ news articles previously clicked by the user. 
Given a candidate news article $v_c \in \mathcal{V}$, the goal of a news recommendation method is to predict the probability $\hat{y}_{u,v_c}$ that user $u$ will click on $v_c$.

\begin{figure*}[tb]
\centering
    \includegraphics[width=0.95\linewidth,trim=25pt 0 25 0]{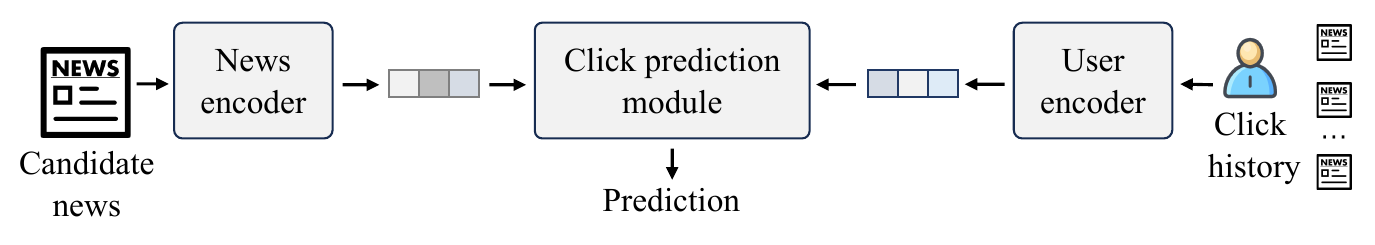}
    \caption{The typical structure of neural news recommendation methods.} 
    \label{fig_intro}
\end{figure*}

A typical (neural) news recommendation method has three components: a news encoder, a user encoder, and a click prediction module, as shown in Fig.~\ref{fig_intro}. 
The news encoder, primarily based on \acp{LM}, is responsible for encoding the textual features $f_t(v)$ and/or other features $f_d(v)$ associated with news article $v$, ultimately producing the news representation $\bm{q}_v$. 
We focus solely on the textual features $f_t(v)$ to compare the ability of different LMs in news modeling within news RSs.
The user encoder generates the user preference representation $\bm{p}_u$ based on the user's click history $H_u$, summarizing the representations of news articles they have browsed.
Using these representations, the click prediction module estimates the click probability $\hat{y}_{u,v_c}$ for candidate news article $v_c$.

\vspace*{-2mm}
\subsection{News recommendation methods}
\vspace*{-1mm}
Following~\citep{plm-nr, speedNRS, miner, efficientFed, tadi23, newsreclib}, we select NAML~\citep{naml}, NRMS~\citep{nrms}, and LSTUR~\citep{lstur} as (neural) news recommendation systems; all involve attention mechanisms for news recommendation. 
In terms of the news encoder, all three use attention for news modeling; NAML incorporates different types of news information, such as titles, bodies, categories, and subcategories, while NRMS focuses solely on learning news representations from titles. 
LSTUR models news representations based on titles and topic categories.
NAML and NRMS employ attention mechanisms to learn user representations, whereas LSTUR uses a GRU network.

As depicted in Fig.~\ref{fig_intro}, the news encoder takes news features as input and then yields the news representation $\bm{q}_v$. 
The user encoder learns the user representation $\bm{p}_u$ based on the user's click history $H_u$, and the click score $\hat{y}$ is computed following the click prediction module $\mathcal{F}^\text{RS} (\cdot)$:
\begin{equation}
    \hat{y}_{u, v} = \mathcal{F}^\text{RS}([\bm{p}_u, \bm{q}_v]).
\end{equation}
For model training, the loss function minimizes the negative log-likelihood of all positive news articles in the ground truth:
\begin{equation}
    \mathcal{L} = -\sum_{u\in \mathcal{U}}\sum_{i\in \mathcal{V}^+} \text{log}\frac{\text{exp}(\hat{y}_{u,i})}{\text{exp}(\hat{y}_{u,i}) + \sum_{j \in \mathcal{V}^-}\text{exp}(\hat{y}_{u,j})}, \label{eq:loss}
\end{equation}
where $\mathcal{V}^+$ is the set of positive new articles for user $u$ in the training dataset, and $\mathcal{V}^-$ is the sampled negative news set corresponding to user $u$ and the $i$-th positive news.
This optimization encourages the model to differentiate between clicked and non-clicked news articles.

\vspace*{-2mm}
\subsection{Language models as news encoders}
\label{sec-LM}

\vspace*{-1mm}
To investigate the effect of different \acp{LM} as news encoders on the performance of news \acp{RS}, we compare three types of \acp{LM} based on model size: \acp{SLM}, \acp{PLM}, and \acp{LLM}.
Each \ac{LM}, when used as a news encoder, can either be used in its \emph{non-fine-tuned} form, relying on its pre-trained knowledge, or it can be \emph{fine-tuned} with additional training on news-specific data to improve the performance on the recommendation task.

\begin{figure}[tb] 
	\centering 
	\subfloat[SLMs in news encoders.]{\label{fig_slm}
		\centering 
		\includegraphics[ width=1\columnwidth,trim=25pt 20 0 10]{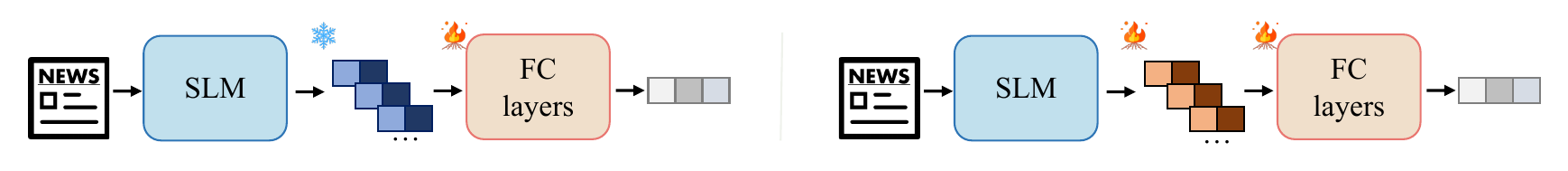}
	}
 \\
	\subfloat[PLMs in news encoders.]{\label{fig_plm}
		\centering
		\includegraphics[width=1\columnwidth,trim=25pt 20 0 10]{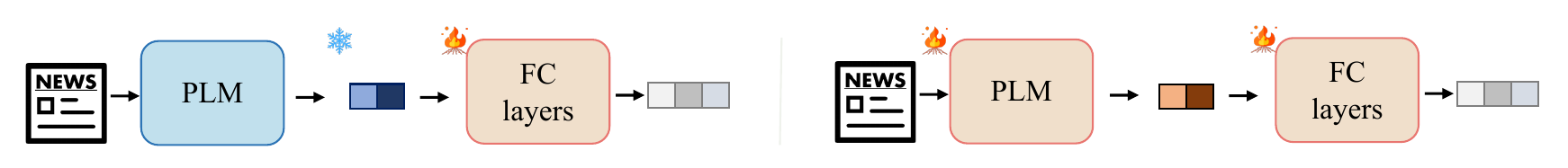}
	}
\caption{SLMs and PLMs as building blocks of news encoders. 
Each LM can be used either in its non-fine-tuned form, shown in the left plots, or in its fine-tuned form, shown in the right plots. The parameters/embeddings in the blue ``ice'' section are fixed, while those in the red ``flame'' section are fine-tuned.
}
\label{fig_lm}
\end{figure}

\header{\acp{SLM} as news encoders} 
Given a news article $v=[v_1, v_2, \ldots, v_n]$, where $v_i$ represents the $i$-th word in article $v$, \acp{SLM} generate static, non-contextualized word embeddings $\bm{e}_{v_i}$ by aggregating global word co-occurrence statistics, with each word having its embedding regardless of context: $\bm{e}_{v_i} = \mathcal{M}^\text{SLM}(v_i; \theta^\text{SLM})$. 
The news representation $\bm{q}_{v}$ is obtained by concatenating the word embeddings of the news content and then applying a fully-connected layer (FC): $\bm{q}_{v} = \text{FC}([\bm{e}_{v_1} \mathbin\Vert \bm{e}_{v_2} \mathbin\Vert \cdots \mathbin\Vert \bm{e}_{v_n}]; \theta^\text{FC})$, where $\mathbin\Vert$ denotes the concatenation operator.\footnote{All FC layers in the paper share a similar architecture, differing primarily in the size of the first layer, which varies depending on the input size to enable the news \ac{RS} to process varying input dimensions.}

\emph{Non-fine-tuned mode.}
As illustrated in the left part of Fig.~\ref{fig_slm}, in the non-fine-tuned mode, \acp{SLM} map each word in a news article to its corresponding embedding $\bm{e}_{v_i}$, which are then concatenated for further processing.
The parameters of the fully connected layer $\theta^\text{FC}$ are tuned according to Eq.~\ref{eq:loss}.

\emph{Fine-tuned mode.}
As illustrated in the right part of Fig.~\ref{fig_slm}, in the fine-tuned mode, the word embeddings $\bm{e}_{v_i}$ and fully-connected layer parameters $\theta^\text{FC}$
are updated based on the recommendation signal (see Eq.~\ref{eq:loss}).

\header{\acp{PLM} as news encoders} 
For a news article $v$, \acp{PLM} first tokenize the news text into  tokens $\mathcal{T}^\text{PLM}(v) = [v_{\text{[CLS]}}^t, v_1^t, \ldots, v_m^t, v_{\text{[SEP]}}^t]$. 
The \acp{PLM} then generate contextualized token embeddings by passing the token sequence through transformer encoder layers: $[\bm{e}_{v_{\text{[CLS]}}^t}, \bm{e}_{v_1^t}, \ldots, \bm{e}_{v_m^t}, \bm{e}_{v_{\text{[SEP]}}^t}] = \mathcal{M}^\text{PLM}([v_{\text{[CLS]}}^t,  v_1^t, \ldots, v_m^t, \\ v_{\text{[SEP]}}^t]; \theta^\text{PLM})$. 
Following~\citep{miner, speedNRS, plm-nr}, we use the embedding of the \texttt{[CLS]} token, which appears at the start of every input sequence, and apply a fully connected layer to represent the entire news article: $\bm{q}_{v} = \text{FC}(\bm{e}_{v_{\text{[CLS]}}^t}; \theta^\text{FC})$.

\emph{Non-fine-tuned mode.}
As shown in the left part of Fig.~\ref{fig_plm}, a \ac{PLM} models the news content and outputs the news embedding. Similar to \acp{SLM}, the parameters of the fully-connected layer $\theta^\text{FC}$ are trained using the loss in Eq.~\ref{eq:loss}.

\emph{Fine-tuned mode.}
As illustrated in the right part of Fig.~\ref{fig_plm}, both the \ac{PLM} parameters $\theta^\text{PLM}$ and the fully-connected layer parameters $\theta^\text{FC}$ are updated during the recommendation process.

\begin{figure}[tb]
\centering
    \includegraphics[width=0.9\linewidth,trim=0 0 0 10]{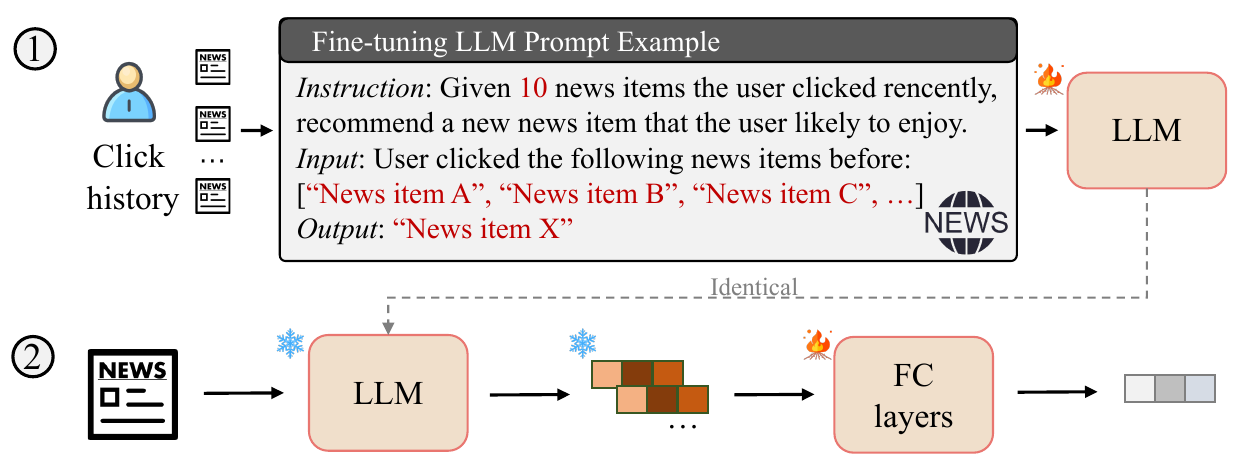}
    \vspace{-4pt}
    \caption{Fine-tuning LLMs as news encoders. In step 1, the LLMs are fine-tuned on news data presented in a natural language format. In step 2, the fine-tuned LLMs generate news embeddings, which are used for the recommendation task.
    } 
    \label{fig_llm}
\end{figure}
\header{\acp{LLM} as news encoders}
For \acp{LLM}, given a news article $v$, we follow the approach in~\citep{llmEmbed}, using fill-in-the-blank prompts, \ie  ``\emph{This news: [$v$] means in one word:}'' to create a prompted version of the news, denoted as $v'$. 
We tokenize it with $\mathcal{T}^\text{LLM}(v') = [{v'}_1^t, {v'}_2^t, \ldots, {v'}_L^t]$.  
Then the \ac{LLM} generates token embeddings: $[\bm{e}_{{v'}_1^t}$, $\bm{e}_{{v'}_2^t}$, \ldots, $\bm{e}_{{v'}_L^t}] = \mathcal{M}^\text{LLM}([{v'}_1^t, {v'}_2^t, \ldots, {v'}_L^t]; \theta^\text{LLM})$.
We use the embeddings of the last $l$ tokens\footnote{$l$ is set to 10 in practice due to computational efficiency.} and apply a fully-connected layer, representing the news article: $\bm{q}_{v} = \text{FC}([\bm{e}_{{v'}_{L-l}^t}, \ldots, \bm{e}_{{v'}_L^t}]; \theta^\text{FC})$.  

\emph{Non-fine-tuned mode.}
Similar to \acp{SLM} and \acp{PLM}, the parameters of the fully-connected layer $\theta^\text{FC}$ are trained in this mode.

\emph{Fine-tuned mode.}
Due to a large number of parameters and high computational costs, as illustrated in Fig.~\ref{fig_llm}, we adopt a two-step process inspired by~\citep{bigRec}. 
First, news recommendation data is transformed into a natural language prompt format, 
and the \ac{LLM} parameters $\theta^\text{LLM}$ are updated using cross-entropy loss to let the \ac{LLM} learn news recommendation-specific information.
Second, the \ac{LLM}, which is fine-tuned in the first step and fixed in the second step, outputs the news embedding for the recommendation process in the same way as in the non-fine-tuned mode, with the fully-connected layer parameters $\theta^\text{FC}$ being updated according to the recommendation objective.

%% file: sections/4-experiments.tex
\section{Experimental Setup}

\vspace*{-2mm}

Below, we detail the dataset and implementation; 
resources to reproduce our results are available at \url{https://github.com/Go0day/LM4newsRec}.
 
\vspace*{-2mm}
\subsection{The MIND dataset}

Following~\citep{nrms, naml, lstur, once24, promptNRS, mtrec22, promptNRSecir}, we conduct experiments using the MIND~\citep{mindDataset} dataset, a public news recommendation dataset collected from the Microsoft News website. 
Table~\ref{tab-mind} provides descriptive statistics for the dataset. 
We use the small version of the original MIND dataset, which is widely adopted in academic research and consists of randomly sampled users and their behavior logs. 
Impressions from November 9 to 14, 2019 are used for training, and those from November 15, 2019 are used for testing~\citep{mindDataset, promptNRSecir}.

\begin{table}[t]
  \centering
  \caption{Statistics of the MIND dataset. } 
  \label{tab-mind}
\begin{tabular}{ wc{1.5cm}  wc{1.5cm}  wc{2.2cm}  wc{2.2cm}  wc{2cm}  wc{2cm}}
\toprule
\#users         & \#news      & \#words in title &  \#words in abs & \#pos clicks & \#neg clicks\\ 
\midrule
94,057          & 65,238      & 11.79    & 38.17           & 347,727     & 8,236,715   \\ 
\bottomrule
\end{tabular}
\end{table}

\vspace*{-2mm}
\subsection{Implementation details}

We use four representative \acp{LM}: 
\textit{GloVe.840B.300d}\footnote{\url{https://nlp.stanford.edu/projects/glove}} (referred to as GloVe), \textit{bert-base-uncased}\footnote{\url{https://huggingface.co/google-bert}} (BERT base version, 110M parameters), \textit{roberta-base}\footnote{\url{https://huggingface.co/FacebookAI}} (RoBERTa, 125M), and \textit{Llama 3.1-8B}\footnote{\url{https://huggingface.co/meta-llama}} (Llama for short). 
These models are selected to cover different families of \acp{LM}. 
To examine the impact of varying model sizes within the same family, we further explore the BERT family by comparing different versions: 
BERT$_{tiny}$ (4.4M parameters), BERT$_{mini}$ (11.3M), BERT$_{small}$ (29.1M), and BERT$_{medium}$ (41.7M).\footnote{\url{https://huggingface.co/prajjwal1}}

Among the selected models, \textit{GloVe.840B.300d} is an \ac{SLM}, \textit{Llama 3.1-8B} an \ac{LLM}, and the rest are \acp{PLM}.
For the \acp{PLM}, we fine-tune varying numbers of layers (from none to all) and select the optimal configuration based on recommendation performance. For Llama, we apply LoRA~\citep{lora} for fine-tuning in step~1, then pre-compute and store news embeddings in advance for recommendation in step~2 (see Fig.~\ref{fig_llm}).
Specifically, NAML~\citep{naml}, NRMS~\citep{nrms}, and LSTUR~\cite{lstur} were originally equipped with GloVe, while PLM-NR~\citep{plm-nr} was originally equipped with the BERT base version. 
We re-implement these foundational publications, standardize them within a unified news recommendation setting (including consistent datasets and model structures), and extend their evaluation by incorporating different LMs.
For all recommendation methods, the maximum length of news titles is set to 20 tokens, and for news abstracts, it is set to 50 tokens; we search the size of the negative clicked news set $|\mathcal{V}^-|$ in Eq.~\ref{eq:loss} from \{1, 2, 3, 4\}, the dropout ratio from \{0.1, 0.2, 0.3, 0.4, 0.5\}, and the learning rate from \{0.0001, 0.00001\}. We use AUC, MRR, nDCG@5~(N@5), and nDCG@10~(N@10) as our evaluation metrics.

\section{Results} 

\vspace*{-2mm}
\subsection{RQ1: Impact of \acp{LM} on news recommendation accuracy}\label{sec:RQ1}
\vspace*{-1mm}
To answer \ref{rq1}, we train news \ac{RS} methods with different sizes of \acp{LM}, as detailed in Section \ref{sec-LM}. The results are reported in Table~\ref{tab-main}. We observe:
\begin{enumerate}[label=(\arabic*)]
    \item GloVe generally yields the lowest performance across different \ac{LM} families, which is expected given its shallow structure. However, it surpasses BERT variants (BERT$_{tiny}$, BERT$_{mini}$, and BERT$_{small}$), showing that larger \acp{LM} do not inherently guarantee superior performance as news encoders.
    \item Comparing BERT and RoBERTa, BERT outperforms RoBERTa in most cases, except on LSTUR. This suggests that BERT may offer more effective news encoding, despite RoBERTa’s higher parameter count.
    \item The performance of Llama does not significantly exceed that of other \acp{LM}, despite its considerably larger parameter count. Thus, an increase in parameters alone does not necessarily translate to better performance.
    \item Within the BERT family, larger models generally achieve better performance than smaller variants. There is one exception: $\text{BERT}_{small}$ does not consistently outperform $\text{BERT}_{mini}$, even with a higher parameter count.
\end{enumerate}
Our findings for \ref{rq1} indicate that, across different \ac{LM} families, larger \acp{LM} do \emph{not} consistently improve news recommendation performance. Within the BERT family, models with more parameters generally perform better; however, this trend is not absolute, as seen in the performance of $\text{BERT}_{mini}$ versus $\text{BERT}_{small}$.

\begin{table}[t!] 
\centering 
\vspace{-5pt}
\caption{Performance comparison of different LMs as news encoders deployed across three news recommendation methods on the MIND dataset. ``BERT'' in the left section denotes BERT base version. Results are averaged over three runs and reported as percentages (\%). Bold font indicates the winner in that column. }\label{tab-main}
\begin{tabular}{c l cccc @{~}@{~}l cccc}
\toprule
    & \multicolumn{5}{c}{Performance}  &  \multicolumn{5}{c}{Performance}
    \\ 
    \cmidrule(r){2-6} \cmidrule{7-11}
    Model & LM      & AUC     & MRR     & N@5   &  N@10 & LM      & AUC     & MRR     & N@5   &  N@10 
    \\ 
    \midrule
    & GloVe   & 66.29 & 31.61 & 34.93  & 41.22    & BERT$_{tiny}$                                   & 64.83 & 30.71 & 33.89  & 40.24 
    \\
    & BERT    & \underline{67.30} & \underline{32.62} & \underline{36.04}  & \underline{42.19}    & BERT$_{mini}$                                   & \underline{65.99} & 31.58 & 34.76  & 41.02 
    \\
    & RoBERTa & 66.73 & 32.10 & 35.52  & 41.64    & BERT$_{small}$                                  & 65.91 & \underline{31.70} & \underline{34.89}  & \underline{41.24} \\ 
\multirow{-4}{*}{\rotatebox{90}{NAML}}  & Llama   & \textbf{68.39} & \textbf{33.20} & \textbf{36.88}  & \textbf{43.06}    & BERT$_{medium}$                                 & \textbf{67.03} & \textbf{32.50} & \textbf{35.97}  & \textbf{42.06} \\ \midrule
                        & GloVe   & 66.62 & 31.34 & 34.83  & 41.04    & BERT$_{tiny}$                                   & 64.30 & 29.10 & 31.76  & 38.56 \\
                        & BERT    & \textbf{68.05} & \textbf{31.80} & \textbf{35.30}  & \textbf{41.72}     & BERT$_{mini}$                                   & \underline{65.70} & \underline{30.34} & \underline{33.32}  & 39.82 \\
                        & RoBERTa & \underline{67.22} & \underline{31.59} & \underline{34.94}  & \underline{41.35}     & BERT$_{small}$                                  &  65.60 & 30.17 & 33.19  & \underline{39.83} \\ 
\multirow{-4}{*}{\rotatebox{90}{NRMS}}  & Llama   & 66.64 & 31.66 & 35.05  & 41.33    & BERT$_{medium}$                                 & \textbf{66.83} & \textbf{31.20} & \textbf{34.49}  & \textbf{40.95} \\ \midrule
                        & GloVe   & 60.43 & 26.26 & 28.62  & 35.11     & BERT$_{tiny}$                                   & 58.48 & 24.47 & 26.41  & 33.09  \\
                        & BERT    & \underline{60.92} & 26.74 & 29.14  & \underline{35.55}    & BERT$_{mini}$                                   & 58.80 & \underline{24.81} & \underline{27.03}  & \underline{33.62} \\
                        & RoBERTa & \textbf{61.25} & \textbf{27.15} & \textbf{29.60}  & \textbf{35.84}    & BERT$_{small}$                                  & \underline{59.14} & 24.65 & 26.66  & 33.55   \\
\multirow{-4}{*}{\rotatebox{90}{LSTUR}} & Llama   & 60.88 & \underline{26.83} & \underline{29.21}  & 35.54    & BERT$_{medium}$                                 & \textbf{59.66} & \textbf{25.39} & \textbf{27.93}  & \textbf{34.38} \\
 \bottomrule                                        
\end{tabular}
\vspace{5pt}
\end{table}

\begin{table}[!ht]  
\centering
\caption{Performance comparison between fine-tuned and non-fine-tuned settings. ``Change'' denotes AUC gain of fine-tuned LMs over non-fine-tuned ones.} \label{tab-finetuned}
\begin{tabular}{c l c @{~}c@{~}@{~}@{~}@{~}@{~}@{~}@{~}c@{~}@{~}@{~}@{~}cc c}
\toprule
\multicolumn{1}{c}{Model}   &   LM        & \multicolumn{1}{c}{Fine-tuned?\ \ \ } & AUC      & MRR      & {nDCG@5}\ \  & {nDCG@10} &{Change}
\\ 
\midrule
                        &  & Y & 66.29    & 31.61    & 34.93    & 41.22    \\
                        & \multirow{-2}{*}{GloVe}   & N  & 65.98    & 31.67    & 35.15    & 41.10    & \multirow{-2}{*}{\textcolor{tablered}{+~0.46\%}} \\ \cmidrule{2-8}
                        &                           & Y          & 67.30    & 32.62    & 36.04    & 42.19    \\
                        & \multirow{-2}{*}{BERT}    & N            & 66.42    & 31.53    & 34.71    & 41.06    & \multirow{-2}{*}{\textcolor{tablered}{+~1.32\%}} \\ 
                        \cmidrule{2-8}
                        &                           & Y          & 66.73    & 32.10    & 35.52    & 41.64    \\
                        & \multirow{-2}{*}{RoBERTa} & N            & 63.51    & 29.25    & 32.29    & 38.70    & \multirow{-2}{*}{\textcolor{tablered}{+~5.07\%}} \\ 
                        \cmidrule{2-8}
                        &                           & Y          & 67.90    & 32.99    & 36.61    & 42.72    \\
\multirow{-8}{*}{\rotatebox{90}{NAML}}  & \multirow{-2}{*}{Llama}   & N            & 68.39    & 33.20    & 36.88    & 43.06    & \multirow{-2}{*}{\textcolor{tableblue}{$-$~0.72\%}} \\ \midrule
                        &                           & Y          & 65.96    & 30.86    & 34.09    & 40.54    \\
                        & \multirow{-2}{*}{GloVe}   & N            & 66.62    & 31.34    & 34.83    & 41.04    & \multirow{-2}{*}{\textcolor{tableblue}{$-$~0.99\%}} \\ \cmidrule{2-8}
                        &                           & Y          & 68.05    & 31.80    & 35.30    & 41.72    \\
                        & \multirow{-2}{*}{BERT}    & N            & 65.70    & 30.56    & 33.33    & 40.13    & \multirow{-2}{*}{\textcolor{tablered}{+~3.58\%}} \\ \cmidrule{2-8}
                        &                           & Y          & 67.22    & 31.59   & 34.94    & 41.35    \\
                        & \multirow{-2}{*}{RoBERTa} & N            & 61.61    & 26.44    & 29.02    & 35.77    & \multirow{-2}{*}{\textcolor{tablered}{+~9.11\%}} \\ \cmidrule{2-8}
                        &                           & Y          & 66.64    & 31.66    & 35.05    & 41.33    \\
\multirow{-8}{*}{\rotatebox{90}{NRMS}}  & \multirow{-2}{*}{Llama}   & N            & 66.56    & 31.71    & 35.13    & 41.30    & \multirow{-2}{*}{\textcolor{tablered}{+~0.11\%}} \\ \midrule
                        &                           & Y          & 60.43    & 26.26    & 28.62    & 35.11    \\
                        & \multirow{-2}{*}{GloVe}   & N            & 58.75    & 25.66    & 27.89    & 34.20    & \multirow{-2}{*}{\textcolor{tablered}{+~2.86\%}} \\ \cmidrule{2-8}
                        &                           & Y          & 60.92    & 26.74    & 29.14    & 35.55    \\
                        & \multirow{-2}{*}{BERT}    & N            & 58.91    & 25.17    & 27.19    & 33.91    & \multirow{-2}{*}{\textcolor{tablered}{+~3.41\%}} \\ \cmidrule{2-8}
                        &                           & Y          & 61.25    & 27.15    & 29.60    & 35.84    \\
                        & \multirow{-2}{*}{RoBERTa} & N            & 57.74    & 24.45    & 26.66    & 32.97    & \multirow{-2}{*}{\textcolor{tablered}{+~6.09\%}} \\ \cmidrule{2-8}
                        &                           & Y          & 60.88    & 26.83    & 29.21    & 35.54    \\
\multirow{-8}{*}{\rotatebox{90}{LSTUR}} & \multirow{-2}{*}{Llama}   & N            & 59.48    & 26.00    & 28.36    & 34.62    & \multirow{-2}{*}{\textcolor{tablered}{+~2.35\%}}
\\ \bottomrule                           
\end{tabular}
\end{table}
\subsection{RQ2: Impact of fine-tuning \acp{LM} on performance and efficiency}

To investigate the role of fine-tuning, we compare the news recommendation accuracy of non-fine-tuned vs.\ fine-tuned models and evaluate the computational efficiency of different \acp{LM}. This section highlights the trade-off between improved performance and computational feasibility.

\begin{figure}[tb]
        \centering
        \subfloat[NAML.]{
            \centering
            \includegraphics[width=0.31\columnwidth]{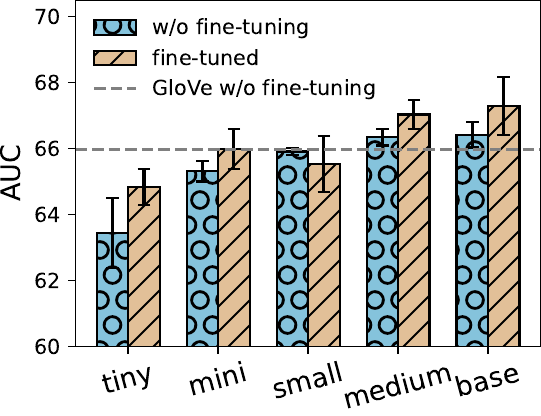}
        }
        \subfloat[NRMS.]{
            \centering
            \includegraphics[width=0.31\columnwidth]{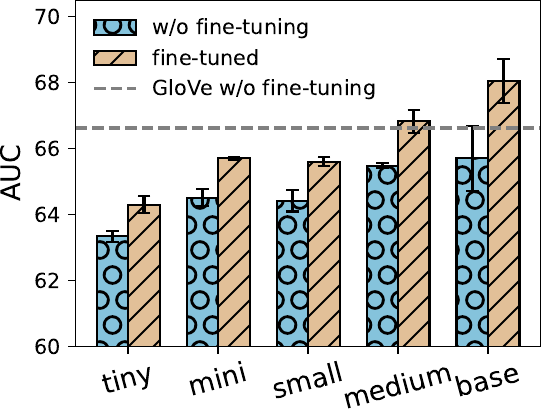}
        }
        \subfloat[LSTUR.]{
            \centering
            \includegraphics[width=0.31\columnwidth]{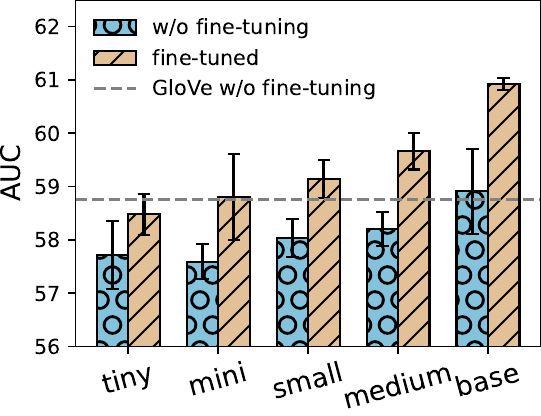}
        }
        \caption{Effect of fine-tuning versus no fine-tuning in the BERT family. } 
        \label{fig-bertfamily}
    \vspace{-0.15cm}
\end{figure}

\smallskip\noindent%
\textbf{Effectiveness.}
Table~\ref{tab-finetuned} and Fig.~\ref{fig-bertfamily} show that fine-tuning \acp{LM} generally improves news recommendation performance, highlighting the effectiveness of fine-tuning. 
However, for Llama, the performance benefits of fine-tuning decline in NAML. 
A plausible reason is that NAML uses both title and abstract text, which may introduce redundancy and noise during Llama’s two-step fine-tuning process.

Within the BERT family (see Fig.~\ref{fig-bertfamily}), we observe that all models benefit from fine-tuning. Notably, in non-fine-tuned settings, most BERT models do not outperform GloVe. This may be due to GloVe’s pre-training on the Common Crawl web data~\cite{glove}, likely making it more suited to news content than BERT models trained on BookCorpus and Wikipedia~\cite{bert}.
The effectiveness of GloVe in representing news content could also explain why fine-tuning leads to a slight performance drop when used in NRMS (see Table~\ref{tab-finetuned}).

\begin{figure}[tb]
    \begin{minipage}[b]{\columnwidth}
        \centering
        \subfloat{
		\includegraphics[width=0.7\textwidth,trim=0 0 0 0]{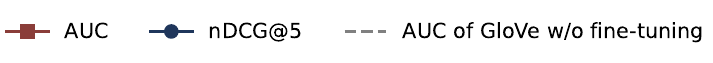}%
	} \setcounter{subfigure}{0}
        \subfloat[NAML.]{
            \centering
            \includegraphics[clip, trim=0pt 0pt 0pt 0, width=0.31\columnwidth]{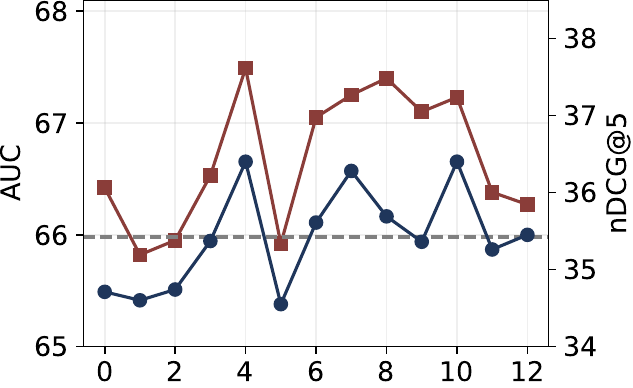}
            \label{layers_bert_naml}
        }
        \subfloat[NRMS.]{
            \centering
            \includegraphics[width=0.31\columnwidth]{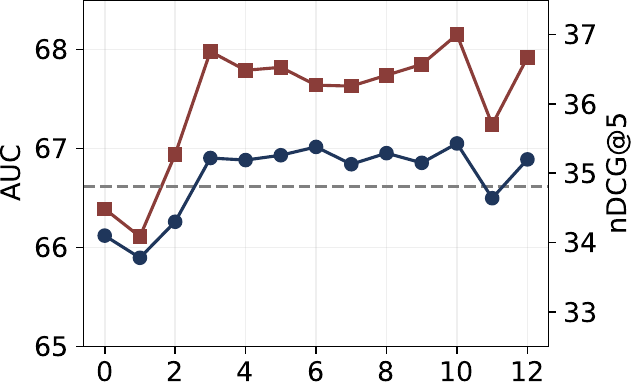}
            \label{layers_bert_nrms}
        }
        \subfloat[LSTUR.]{
            \centering
            \includegraphics[width=0.31\columnwidth]{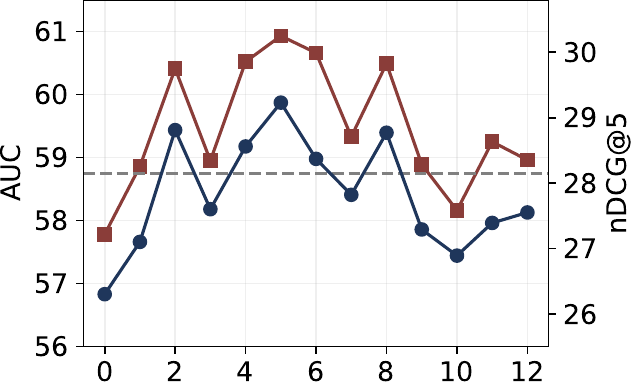}
            \label{layers_bert_lstur}
        }
        \caption{Effect of varying the number of fine-tuned layers in BERT.} 
        \label{fig-layers}
    \end{minipage}
    \vspace{-10pt}
\end{figure}

Additionally, when analyzing BERT (base) by fine-tuning different numbers of layers as shown in Fig.~\ref{fig-layers}, we observe that the optimal number of layers varies significantly across different RS methods.
Interestingly, in some cases, fine-tuned BERT does not outperform non-fine-tuned GloVe, further underscoring GloVe's strength in capturing news representations.

\begin{figure}[t]
\centering
    \includegraphics[width=1.0\linewidth]{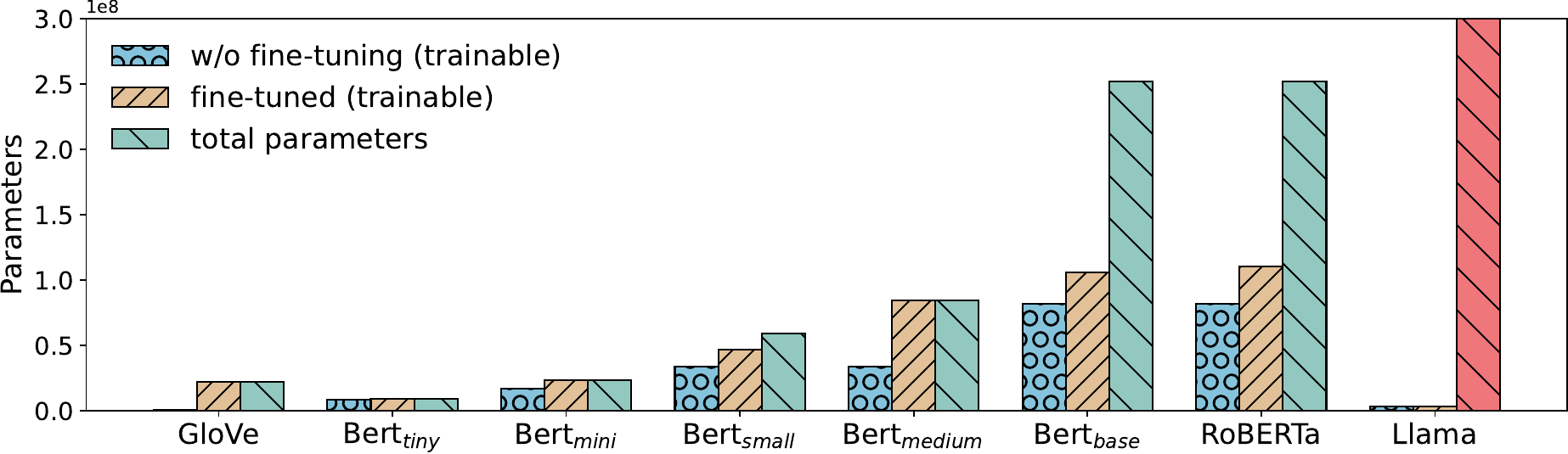}
    \caption{Comparison of trainable and total parameters for different LMs in fine-tuned and non-fine-tuned modes within the NAML framework. Llama's total parameter (over 8 billion) is highlighted in red as it significantly exceeds the scale of other models and cannot be visually included within the same figure.
    } 
    \label{fig_parameters}
\end{figure}

\smallskip\noindent%
\textbf{Efficiency.}
Fig.~\ref{fig_parameters} provides parameter statistics within the NAML framework, with similar trends in NRMS and LSTUR. ``Total parameters'' represents all parameters involved in the recommendation process, while ``trainable parameters'' includes only those updated during training (see Section \ref{sec-LM}). 
Generally, as \ac{LM} size increases, both total and trainable parameters grow. However, for Llama, the two-step fine-tuning strategy and the use of pre-computed news embeddings significantly reduce its trainable parameters. 
And for GloVe, the concatenation of word embeddings results in a higher number of trainable parameters than BERT$_{tiny}$ and is comparable to BERT$_{mini}$.

Overall, these findings underscore different trade-offs when selecting a \ac{LM} as news encoder (answering \ref{rq2}): GloVe is an efficient option for cases with limited fine-tuning resources, offering effective performance with minimal computational demands. For static datasets without frequent news updates, precomputing Llama embeddings and using them for inference offers a high-performance alternative. Finally, when both computational resources and performance are priorities, fine-tuning the BERT base version for the news recommendation task provides a balanced, high-performing solution.

\vspace*{-2mm}
\subsection{RQ3: Impact of \acp{LM} on cold-start user performance}
\begin{figure}[tb]
    \begin{minipage}[b]{\columnwidth} 
        \centering
        \subfloat{
		\includegraphics[width=0.9\textwidth,trim=0 0 0 0]{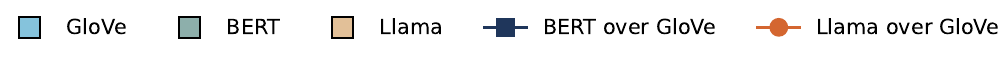}
	} \setcounter{subfigure}{0}
        \subfloat[NAML.]{
            \centering
            \includegraphics[clip, trim=0pt 0pt 0pt 0, width=0.3285\columnwidth]{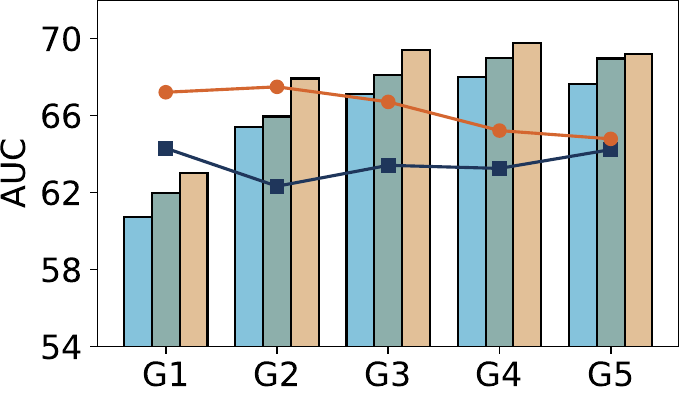}
        }
        \subfloat[NRMS.]{
            \centering
            \includegraphics[width=0.282\columnwidth]{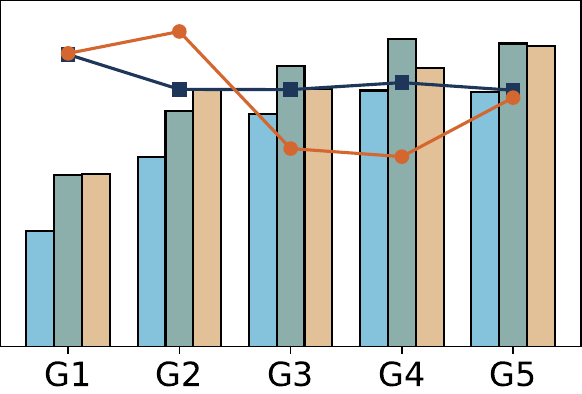}
        }
        \subfloat[LSTUR.]{
            \centering
            \includegraphics[width=0.339\columnwidth]{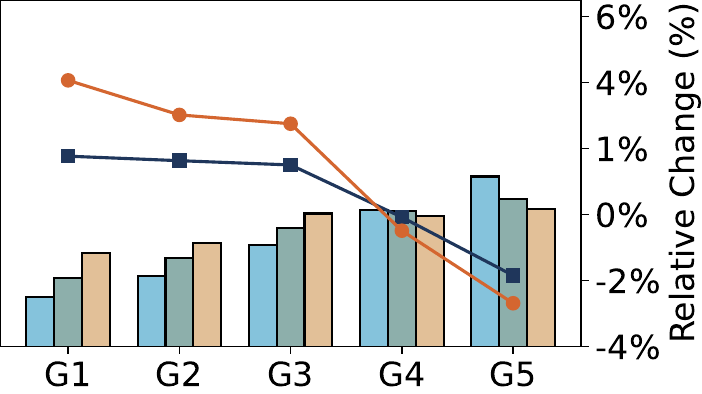}
            \label{fig-cold-lstur}
        }
        \caption{Effect of \acp{LM} across user groups with varying click history lengths. User groups `1' through `5' represent progressively longer click histories. The `Relative Change' indicates each \ac{LM}'s performance improvement ratio compared to GloVe.} 
        \label{fig-cold}
    \end{minipage}
   \vspace{0.15cm}
\end{figure}

Given the inconsistent results regarding performance gains with larger \acp{LM} as news encoders (see Section \ref{sec:RQ1}), we further investigate whether larger \acp{LM} benefit specific user groups in news \acp{RS}.
To examine this, we test three representative \acp{LM}: GloVe, BERT (base), and Llama. 
The users are sorted by click history length and categorized into five engagement levels: Group 1 (0--20\%), Group 2 (20\%--40\%), Group 3 (40\%--60\%), Group 4 (60\%--80\%), and Group 5 (80\%--100\%), representing different levels of engagement based on the distribution of click history length.
The average click lengths for each group, \ie the number of news articles previously clicked by users in each group, are 4.01, 9.27, 16.57, 29.60, and 48.58, respectively. 
In Fig.~\ref{fig-cold}, the bars show the AUC scores for various \acp{LM} as news encoders across user groups, while the line plot illustrates each \ac{LM}'s relative change over GloVe.

We find that Llama provides the greatest improvement in Group 1, which includes the ``coldest'' users with the smallest amount of click history.
This improvement may be due to the limited interaction data available for these users, where larger \acp{LM} can leverage richer text-based representations to alleviate sparse click signals.
As users’ click history expands (\eg Group 5), the relative benefit of larger \acp{LM} diminishes, indicating that user engagement itself provides a strong signal for modeling. 
Interestingly, in the LSTUR model~(see Fig.~\ref{fig-cold-lstur}), the relative improvement from larger \acp{LM} decreases more sharply and even turns negative in Group 5. 
This may be attributed to LSTUR's use of GRU for modeling user preferences, which, unlike the attention mechanisms used in NAML and NRMS, is less effective at capturing evolving user interests~\cite{sasrec}. 
As news representations become more comprehensive with larger \acp{LM}, the limitations in modeling dynamic user preferences likely contribute to the observed performance declines.

In response to \ref{rq3}, these findings suggest that larger \acp{LM} enhance performance for cold-start users. 
Their effectiveness decreases as click history increases, especially in news \acp{RS} with limited user modeling capabilities.

%% file: sections/5-limitations.tex
\section{Limitations and Broader Impact}
Our study has several limitations. First, we only use the MIND-small dataset for news recommendation due to resource constraints.
Additionally, we limit our analysis to English news datasets to focus on evaluating model size effectiveness, leaving out datasets in other languages, such as EB-NeRD~\cite{ebnerd}, Adressa~\cite{adressa}, and Plista~\cite{plista}. Investigating the impact of \acp{LM} in non-English news recommendation would be an interesting and valuable direction.
Second, we examine \acp{LM} with a maximum of 8 billion parameters (Llama) as news encoders, as evaluating larger models (e.g., 13 billion, 70 billion, etc.) exceeded our resource capacity. We expect that larger models might offer further gains, particularly for cold-start users.
Third, our study explores three news recommendation methods: NAML, NRMS, and LSTUR, which are commonly used as benchmarks~\cite[see, e.g.,][]{plm-nr, efficientFed, surveyWuNews}. 
In news recommendation, a significant proportion of news articles that are awaiting recommendations do not appear in the logged data. Specifically, in the MIND dataset we used, approximately 32.9\% of the news articles in the test set never appear in the training set. This makes ID-based collaborative filtering methods, such as matrix factorization and graph-based approaches, unsuitable for our setting. Therefore, we focus on these three representative content-based news recommendation methods and leave the exploration of other techniques for future work.

Beyond limitations, our study has broader impacts. It provides a reference point for both academia and industry regarding the role of \acp{LM} in news recommendation, showing that larger models do not always translate to better performance. Our findings demonstrate that deploying \acp{LM} can help address the cold-start problem for new users, enhancing recommendation reliability for underrepresented groups. 
We believe our work has the potential to contribute to advancing socially responsible and reliable news recommendation systems.

%% file: sections/6-conclusion.tex
\vspace*{-2mm}
\section{Conclusion}

\vspace*{-1mm}
In this work, we have revisited the role of \acfp{LM} as news encoders within neural news \acp{RS} on the MIND dataset. 
We have investigated the effects of varying \ac{LM} sizes, assessing the impact of fine-tuning on recommendation performance and analyzing model performance across different user groups.

Our main finding is that larger \acp{LM} as news encoders do not consistently yield better recommendation results, contrasting with previous studies~\cite{plm-nr, efficientFed}. Additionally, we observe that larger \acp{LM} require more precise fine-tuning and greater computational resources, prompting a trade-off consideration based on performance needs and resource availability.

Notably, we identify an interesting tendency: larger \acp{LM} show more significant improvements in recommendations for cold-start users, suggesting potential benefits in modeling user interests with limited click history. 
A promising future direction is to investigate the stability of \ac{LM}-based \acp{RS} as the news \ac{RS} domain evolves. Additionally, exploring the design of larger \acp{LM} to better meet the dynamic needs of diverse user groups would be valuable.